\documentclass[aps, prd, amsmath, floats, floatfix, onecolumn, nofootinbib, superscriptaddress, showpacs]{revtex4-1}
\usepackage{graphicx}
\usepackage{color}
\usepackage{latexsym}
\usepackage{amsfonts}
\usepackage[colorlinks]{hyperref}

\newcommand{\be}{\begin{eqnarray}}
\newcommand{\ee}{\end{eqnarray}}

\begin{document}

\title{Boosted Kerr black hole in the presence of plasma\footnote{Presented at the meeting {\it Recent Progress in Relativistic Astrophysics}, 6-8 May 2019 (Shanghai, China).}}

\author{Carlos A. Benavides-Gallego}
\email[Corresponding author: ]{abgcarlos17@fudan.edu.cn}
\affiliation{Center for Field Theory and Particle Physics and Department of Physics, Fudan University, 200433 Shanghai, China}

\author{Ahmadjon~Abdujabbarov}
\email[Corresponding author: ]{ahmadjon@astrin.uz}
\affiliation{Center for Field Theory and Particle Physics and Department of Physics, Fudan University, 200433 Shanghai, China}
\affiliation{ Ulugh Beg Astronomical Institute, Astronomicheskaya 33, Tashkent  100052, Uzbekistan; ahmadjon@astrin.uz}
%\firstnote{Presented at the meeting {\it Recent Progress in Relativistic Astrophysics}, 6-8 May 2019 (Shanghai, China).}

\date{\today}

\begin{abstract}
In this work, we obtain the deflection angle for a boosted Kerr black hole in the weak field approximation using the optics in a curved spacetime developed by J.~L.~Synge in 1960. We study the behavior of light in the presence of plasma by considering different distributions: uniform plasma, singular isothermal sphere, non-singular isothermal gas sphere, and plasma in a galaxy cluster. We found that the dragging of the inertial system along with the boosted parameter $\Lambda$ affect the value of the deflection angle. As an application, we studied the magnification for both uniform and singular isothermal distributions.
\end{abstract}

\maketitle

%%%%%%%%%%%%%%%%%%%%%%%%%%%%%%%
\section{Introduction}
The detection of gravitational waves from the coalescence of two black holes showed the formation of a rapidly rotating black hole boosted with linear velocity~\cite{LIGO16b, LIGO16c, LIGO16d}. Furthermore, the possible observation of the electromagnetic counterpart from black hole merger could provide more information about angular and linear momentum in such systems~\cite{Morozova14, Lyutikov11}. In this sense, it is important to include the boost parameter into the Kerr black hole solution to investigate possibles effects on the gravitational field.

On the other hand, gravitational lensing has been used as a tool to test Einstein's theory of gravity and a lot of work has been done in this field by the scientific community. Starting from the work of Synge on optics in a curved spacetime~\cite{Synge60}, for example, it has been shown that the photon motion is affected by the presence of plasma~\cite{Kogan10,Tsupko12,Morozova13}. Moreover, the effect of plasma around a compact objects on lensing effects has been studied in~\cite{Rogers:2017ofq,Rogers:2016xcc,Kichenassamy:1985zz,Perlick:2017fio,Perlick:2015vta} and references there in. In the literature, there is also a lot of work devoted to the so-called black hole shadow~\cite{Vries1999,Abdujabbarov:2017pfw,Abdujabbarov:2016hnw,Abdujabbarov:2015xqa,Abdujabbarov:2015pqp,Bambi:2008jg,Bambi:2010hf,Li:2013jra} and references there in. 

Recently,  a solution of Einstein's vacuum field equations which describes a boosted Kerr black hole relative to an asymptotic Lorentz frame was obtained in Ref.~\cite{Soares17}. This solution opens the possibility to study the weak gravitational lensing effect around a boosted Kerr black hole in the presence of plasma. Hence, in the present manuscript, we study the gravitational lensing effect around a boosted black hole by considering different plasma distributions. The paper is organized as follow. First, we briefly discuss the optics in curved spacetimes and describe the procedure to obtain the deflection angle in the weak field approximation following works of Ref.~\cite{Kogan10, Morozova13}. Then, we present the boosted Kerr metric in both non-rotating and slowly rotating cases. Next, following the previous discussion, we find the deflection angle and study its behavior in the presence of plasma, both for uniform and non-uniform distributions (singular isothermal sphere (\textbf{SIS}), non-singular isothermal sphere (\textbf{NSIS}), and the case of a plasma in a galaxy cluster (\textbf{PGC}).) Finally, as an application, we study the magnification for uniform and \textbf{SIS} plasma distributions.

Throughout the paper, we use the convention in which Greek indices run from 0 to 3, while Latin indices run from 1 to 3. Moreover, we use geometrized units, where $c=G=1$. This work was presented as a talk at the meeting: \textit{Progress in Relativistic Astrophysics} at Fudan University and it is based on the paper: "Gravitational lensing for a boosted Kerr black hole in the presence of plasmas" published by the European physics journal C.   
\section{Optics in a curved space-time \label{sec:Optics} }
The optics in a curved space-time was first developed by Synge in 1960~\cite{Synge60}. This approach considers a static spacetime metric which describes a weak gravitational field in an asymptotically flat spacetime. In this sense, the metric tensor $g_{\mu\nu}$ can be expressed as 
\begin{equation}
\label{I.1}
\begin{array}{ccc}
g^{\alpha\beta}=\eta^{\alpha\beta}-h^{\alpha\beta}&\text{and}&g_{\alpha\beta}=\eta_{\alpha\beta}+h_{\alpha\beta}.
\end{array}
\end{equation}
Here $\eta_{\alpha\beta}$ is the Minkowski spacetime metric tensor and $h_{\alpha,\beta}$ are small perturbation ($h_{\alpha\beta}\ll 1$). Moreover, because the metric must be asymptotically flat, the small perturbations satisfies $h_{\alpha\beta}\rightarrow 0$ for $x^\alpha \rightarrow \infty$ and $h^{\alpha\beta}=h_{\alpha\beta}$.

Our main interest is to study the effect of different plasma distributions on the gravitational lensing for a boosted Kerr black hole. In this sense, it is necessary to obtain the photon trajectories in the presence of a gravitational field and then include the plasma contribution into the equations of motion. To do so, Synge modified the Fermat's least action principle for light propagation by considering a dispersive medium. Hence, using the Hamiltonian formalism, he was able to show that the variational principle
\begin{equation}
\label{I.2}
\delta\left(\int p_\alpha dx^\alpha \right)=0,
\end{equation}
along with the condition 
\begin{equation}
\label{I.3}
W(x^\alpha,p_\alpha)=\frac{1}{2}\left[g^{\alpha\beta}p_\alpha p_\beta-(n^2-1)\left(p_0\sqrt{-g^{00}}\right)^2\right]=0
\end{equation}
leads to the following system of differential equations~\cite{Synge60}
\begin{equation}
\label{I.4}
\begin{array}{ccc}
\frac{dx^\alpha}{d\lambda}=\frac{\partial W}{\partial p_\alpha}&\text{and}&\frac{dp_\alpha}{d\lambda}=-\frac{\partial W}{\partial x^\alpha},
\end{array}
\end{equation}
where the affine parameter $\lambda$ changes along the light trajectory. These equations describe the trajectories of photons in a gravitational field. On the other hand, it is important to point out that the scalar function $W(x_\alpha, p_\alpha)$ has been defined from the relationship between the phase velocity\footnote{The phase velocity is defined as the minimum value of
	\begin{equation*}
	u'^2=1+\frac{dx_\alpha dx^\alpha}{(V_\beta dx^\beta)^2},
	\end{equation*}	
	where $u'$ is the velocity of a fictitious particle riding on the wavefront relative to a time-like world-line $C$ (intersecting the wave) of an observer with 4-velocity $V^\mu$ (see \cite{Synge60} for details).} $u$ and the 4-vector of the photon momentum $p^\alpha$, which is given by~\cite{Synge60}
\begin{equation}
\label{I.5}
\frac{c^2}{u^2}=n^2=1+\frac{p_\alpha p^\alpha}{(p^0\sqrt{-g_{00}})^2}.
\end{equation}
Now, to include the effect of plasma in the equations of motion for photons, we can consider a static inhomogeneous plasma with a refraction index $n$ that depends on the space location $x^i$. Mathematically, this refraction index is given by~\cite{Kogan10,Morozova13}
\begin{equation}
\label{I.6}
\begin{array}{ccc}
n^2=1-\frac{\omega_e^2}{[\omega(x^i)]^2} &\text{and}&\omega^2_e=\frac{4\pi e^2 N(x^i)}{m}=K_eN(x^i).
\end{array}
\end{equation} 
In the last expression, $e$ and $m$ are the electron charge and mass respectively, $\omega_e$ is the plasma frequency, and $N(x^i)$ is the electron concentration in an inhomogeneous plasma. Moreover, the photon frequency $\omega(x^i)$ depends on the space coordinates $x^1$, $x^2$, $x^3$ due to gravitational redshift. 

It is known that for a static medium in a static gravitational field, the photon energy can be expressed as~\cite{Synge60}
\begin{equation}
\label{I.7}
p_0\sqrt{-g^{00}}=-\frac{1}{c}\hbar\omega(x^i).
\end{equation} 
Therefore, after using Eq.~(\ref{I.6}) the scalar function $W(x^\alpha,p_\alpha)$ reduces to
\begin{equation}
\label{I.8}
W(x^\alpha,p_\alpha)=\frac{1}{2}\left[g^{\alpha\beta}p_{\alpha}p_{\beta}+\frac{\omega^2_e\hbar^2}{c^2}\right],
\end{equation}
where $\hbar$ is the Planck's constant. The scalar function expressed in Eq.~(\ref{I.8}) has been used in Refs.~\cite{Kogan10,Morozova13} to find the equations of light propagation for diagonal and non-diagonal spacetimes.

In contrast to flat spacetime in vacuum, where the trajectories of photons are straight lines, the presence of an arbitrary medium in curved spacetimes makes photons move along bent trajectories. However, taking into account only small deviations, it is possible to use the components of the 4-momentum of the photon moving in a straight line along the $z-$axis as an approximation. This components are given by~\cite{Kogan10,Morozova13} 
\begin{equation}
\label{I.9}
\begin{array}{ccc}
p^\alpha=\left(\frac{\hbar\omega}{c},0,0,\frac{n\hbar\omega}{c}\right)&\text{and}&p_\alpha=\left(-\frac{\hbar\omega}{c},0,0,\frac{n\hbar\omega}{c}\right).
\end{array}
\end{equation}
Eqs.~(\ref{I.9}) are known as the null approximation. It is important to point out that both $\omega$ and $n$ are evaluated at $\infty$. In this sense, we have introduced the notation in which~\cite{Kogan10,Morozova13}
\begin{equation}
\begin{array}{ccc}
\omega=\omega(\infty)&\text{and}&n=n(\infty).\\
\end{array}
\end{equation}             
Now, in sections \ref{sec:I.1} and \ref{sec:I.2}, we use the equations of motion to compute the deflection angle for diagonal and non-diagonal spacetimes. We shall see that it depends on the small perturbations $h_{ij}$ and the plasma distribution.
\subsection{Equations of light propagation in a diagonal spacetime \label{sec:I.1}}
In a diagonal spacetime, the non-zero components of metric tensor  $g_{\alpha\beta}$ are those with $\alpha=\beta$. Hence, in this particular case, the function $W(x^\alpha,p_\alpha)$ takes the form
\begin{equation}
\label{2.1.1}
W(x^\alpha,p_\alpha)=\frac{1}{2}\left[g^{00}p^2_0+g^{lm}p_{l}p_{m}+\frac{\omega^2_e\hbar^2}{c^2}\right],
\end{equation}
and the equations of motion for photos in the presence of plasma (see Eq.~(\ref{I.4})) reduces to 
\begin{equation}
\label{2.1.2}
\begin{aligned}
\frac{dx^i}{d\lambda}&=g^{ij}p_j\\
\frac{dp_i}{d\lambda}&=-\frac{1}{2}g^{lm}_{\;\;\;\;,i} p_l p_m-\frac{1}{2}g^{00}_{\;\;\;\;,i}p^2_0-\frac{1}{2}\frac{\hbar^2}{c^2}K_eN_{,i}.
\end{aligned}
\end{equation}
Now, due to the null approximation, the 3-vector in the direction of the photon's momentum (first equation in Eq.~(\ref{2.1.2})) can be expressed as
\begin{equation}
\label{2.1.3}
p_i=\frac{n\hbar\omega}{c}e_i,
\end{equation} 
where $e_i=(0,0,1)$. Therefore, the second relation in Eq.~(\ref{2.1.2}) takes the form

\begin{equation}
\label{2.1.4}
\frac{d}{d\lambda}\left(\frac{n\hbar\omega}{c}e_i\right)=-\frac{1}{2}g^{lm}_{\;\;\;\;,i}p_lp_m\nonumber-\frac{1}{2}g^{00}_{\;\;\;\;,i}p^2_0-\frac{1}{2}\frac{\hbar^2}{c^2}K_eN_{,i}.
\end{equation}
The last expression can be written as 

\begin{equation}
\label{2.1.5}
\frac{de_i}{dz}=-\frac{1}{2}\frac{c^2}{n\hbar^2\omega^2}\left(g^{00}_{\;\;\;\;,i}(p_0)^2+g^{lm}_{\;\;\;\;,i}p_l p_m+\frac{\hbar^2}{c^2}K_eN_{,i}\right)\nonumber\\
-e_i\frac{dn}{dz}.
\end{equation}
According to reference~\cite{Kogan10}, only the components of $e_i$ that are perpendicular to the initial direction of propagation were taken into account. This means that the contribution to the deflection of photons is due only to the change in $e_1$ and $e_2$. Hence, after using the null approximation $e_i=0$ along with the assumption of a weak gravitational field, the last equation reduces to 
\begin{equation}
\label{2.1.6}
\frac{de_i}{dz}=\frac{1}{2}\left(h_{33,i}+\frac{1}{n^2}h_{00,i}-\frac{1}{n^2\omega^2}K_eN_{,i}\right),
\end{equation}   
for $i=1,2$. Equation (\ref{2.1.6}) can be used to obtain the deflection angle, which is defined by~\cite{Perlick00,Perlick:2004tq,Schneider92} 
\begin{equation}
\label{2.1.7}
\vec{\hat{\alpha}}=\mathbf{e}(+\infty)-\mathbf{e}(-\infty).
\end{equation}
Thus, after integration, we obtain the following expression for $\hat{\alpha}$
\begin{equation}
\label{2.1.8}
\hat{\alpha}_i=\frac{1}{2}\int^{\infty}_{-\infty}\left(h_{33,i}+\frac{\omega^2}{\omega^2-\omega^2_e}h_{00,i}-\frac{K_e}{\omega^2-\omega^2_e}N_{,i}\right)dz,
\end{equation}
for $i=1,2$. Note that $\omega_e$ and $n$ are evaluated at infinty. Finally, in terms of the impact parameter, $\hat{\alpha}_i$ can be expressed as
\begin{equation}
\label{2.1.9}
\hat{\alpha}_b=\frac{1}{2}\int^\infty_{-\infty}\frac{b}{r}\left(\frac{dh_{33}}{dr}+\frac{1}{1-\omega^2_e/\omega^2}\frac{dh_{00}}{dr}-\frac{K_e}{\omega^2-\omega^2_e}\frac{dN}{dr}\right).
\end{equation}

\subsection{Equations of light propagation in a non-diagonal spacetime \label{sec:I.2}}
In a non-diagonal spacetime the components of the metric tensor $g_{\alpha\beta}$ do not vanish for $\alpha\neq\beta$. In this sense, the escalr function $W(x^\alpha,p_\alpha)$ in a non-diagonal spacetime takes the form
\begin{equation}
\label{2.2.1}
W(x^\alpha,p_\alpha)=\frac{1}{2}\left[g^{00}p^2_0+2g^{0l}p_{0}p_{l}+g^{lm}p_{l}p_{m}+\frac{\omega^2_e\hbar^2}{c^2}\right].
\end{equation}
Note that Eq.~(\ref{2.2.1}) differs from Eq.~(\ref{2.1.1}) only in terms of the form $g^{0l}p_{0}p_{l}$. Therefore, after substitution, the equations of motion for photons in the presence of plasma reduces to 
\begin{equation}
\label{2.2.2}
\begin{aligned}
\frac{dx^i}{d\lambda}&=g^{ij}p_j\\
\frac{dp_i}{d\lambda}&=-\frac{1}{2}g^{lm}_{\;\;\;\;,i}p_lp_m-\frac{1}{2}g^{00}_{\;\;\;\;,i}p^2_0-g^{0l}_{\;\;\;\;,i}p_0p_l-\frac{1}{2}\frac{\hbar^2}{c^2}K_eN_{,i}.
\end{aligned}
\end{equation}
Once again, the first equation in Eq.~(\ref{2.2.2}) can be expressed as Eq.~(\ref{2.1.3}). Hence, after considering the null and the weak field approximation, the second equation in Eq.~(\ref{2.2.2}) can be expressed as
\begin{equation}
\label{2.2.3}
\frac{dp_i}{dz}=\frac{1}{2}\frac{n\hbar\omega}{c}\left(h_{33,i}+\frac{1}{n^2}h_{00,i}+\frac{1}{n}h_{03,i}-\frac{K_eN_{,i}}{n^2\omega^2}\right),
\end{equation}
from which, after integration, the deflection angle for a non-diagonal spacetime in the presence of plasma is given by 
\begin{equation}
\label{2.2.4}
\hat{\alpha}_i=\frac{1}{2}\int^{\infty}_{-\infty}\bigg(h_{33,i}+\frac{\omega^2}{\omega^2-\omega^2_e}h_{00,i}+\frac{1}{n}h_{03,i}-\frac{K_eN_{,i}}{\omega^2-\omega^2_e}\bigg)dz\ .
\end{equation}
\section{Boosted Kerr black hole in the presence of plasma}
The boosted Kerr black hole was obtained by I.~D.~Soares in 2017~\cite{Soares17}. This spacetime is a solution of Einstein's vacuum field equations which describe a boosted black hole relative to an asymptotic Lorentz frame. In the Kerr-Schild coordinates, the line element is given by 
\begin{equation}
\label{3.1}
\begin{aligned}
ds^2&=-\left(1-\frac{2Mr}{\Sigma}\right)dt'^2+\left(1+\frac{2Mr}{\Sigma}\right)dr^2\nonumber+\frac{\Sigma}{\Lambda}d\theta^2+\frac{A\sin^2(\theta)}{\Lambda^2\Sigma}d\phi^2\\\\
&-\frac{4Mra\sin^2\theta}{\Lambda\Sigma}dt'd\phi-\frac{4Mr}{\Sigma}dt'dr-\frac{2a\sin^2\theta}{\Lambda}\left(1-\frac{2Mr}{\Sigma}\right)drd\phi
\end{aligned}
\end{equation} 
with,
\begin{equation}
\label{3.2}
\begin{array}{ccc}
\Sigma=r^2+a^2\left(\frac{\beta+\alpha\cos\theta}{\alpha+\beta\cos\theta}\right)^2,&\Lambda=(\alpha+\beta\cos\theta)^2,&A=\Sigma^2+a^2\left(\Sigma+2Mr\right)\sin^2\theta.
\end{array}
\end{equation}
Note that the solution has three parameters: mass, rotation, and boots. Moreover, $a={J}/{M}$ is the specific angular momentum of the compact object with total mass $M$, $\alpha=\cosh\gamma$, $\beta=\sinh\gamma$, and $\gamma$ is the usual Lorentz factor which defines the boost velocity $v$ by the formula $v=\tanh\gamma={\beta}/{\alpha}$. The metric in~(\ref{3.1})
exactly reduces to Kerr when $\Lambda=1$ ($v=0$). Moreover, it is important to point out that the direction of the boost for the Kerr black hole is along the axis of rotation.

In this section, we compute the deflection angle for a boosted Kerr black hole in the presence of plasma using the weak field approximation discussed in Sec.\ref{sec:Optics}. First, we consider the non-rotating case ($a=0$) which correspond to the boosted Schwarzschild black hole, and then the slowly rotating case.

%%%%%%%%%%%%%%%%%%%%%%%%%%%%%%%%%%%%%%%%%%%%%%%%%%%%%%%%%%%%%

\subsection{Non-rotating case}
To study the behavior of $\hat{\alpha}$ in the presence of plasma for the non-rotating case, it is necessary to express the line element Eq.~(\ref{3.1}) in Cartesian coordinates and find the small perturbations $h_{ij}$. Nevertheless, before doing so, we first consider the limit $v<<1$. Hence, under this approximation, Eq.~(\ref{3.1}) takes the form~\cite{Benavides-Gallego:2018ufb}
\begin{equation}
\label{3.1.1}
\begin{aligned}
ds^2&=-\left(1-\frac{2M}{r}\right)dt'^2+\left(1+\frac{2M}{r}\right)dr^2+r^2(1-2v\cos\theta)d\theta^2+r^2\sin^2\theta d\phi^2\nonumber\\
&-4vr^2\sin^2\theta\cos\theta d\phi^2-\frac{4M}{r}dt'dr.
\end{aligned}
\end{equation}
Now, using the coordinate transformation
\begin{equation}
\label{3.1.2}
\begin{array}{cccc}
\overline{t}=t,&\overline{x}=r\sin\theta\cos\phi,&\overline{y}=r\sin\theta\sin\phi,&\overline{z}=r\cos\theta.
\end{array}
\end{equation}
Eq.~(\ref{3.1.1}) reduces to 
\begin{equation}
\begin{aligned}
\label{3.1.3}
ds^2&=ds^2_0+h_{11}d\overline{x}^2+h_{12}d\overline{x}d\overline{y}+h_{13}d\overline{x}d\overline{z}+h_{22}d\overline{y}^2+h_{23}d\overline{y}d\overline{z}+h_{00}dt^2+d\overline{z}^2h_{33},
\end{aligned}
\end{equation}
from which,
\begin{equation}
\label{3.1.4}
\begin{array}{ccc}
h_{00}=\frac{2M}{r}&\text{and}&h_{33}=\frac{2M}{r}\cos^2\theta-2v\cos\theta\sin^2\theta.
\end{array}
\end{equation}
The expressions for $h_{11}$, $h_{12}$, $h_{13}$, $h_{22}$, and $h_{23}$ can be found in ref.~\cite{Benavides-Gallego:2018ufb}. Here $ds^2_0=-dt^2+d\overline{x}^2+d\overline{y}^2+d\overline{z}^2$. Now, after using~Eq.(\ref{2.1.9}), the deflection angle in the non-ratating case is given by
\begin{equation}
\label{3.1.5}
\hat{\alpha}_b=\frac{2M}{b}+\frac{2Mb}{1-\frac{\omega^2_e}{\omega^2}}\int^\infty_{0}\frac{dz}{(b^2+z^2)^\frac{3}{2}}+\frac{bK_e}{2}\int^\infty_{-\infty}\frac{1}{\omega^2-\omega^2_e}\frac{1}{r}\frac{dN}{dr}dz.
\end{equation}
From Eq.~(\ref{3.1.5}) we note that, at first order, $\hat{\alpha}_b$ does not depend on the velocity. Hence, if we consider a uniform plasma ($\omega_e$ constant), and the approximation $1-n\ll\frac{\omega_e}{\omega}$, Eq.~(\ref{3.1.5}) reduces to~\cite{Kogan10}
\begin{eqnarray}
\label{3.1.6}
\hat{\alpha}_b=\frac{2M}{b}\left(1+\frac{1}{1-\frac{\omega^2_e}{\omega^2}}\right).
\end{eqnarray} 

\begin{figure}[h!]
	\centering	
	a.\includegraphics[scale=0.34]{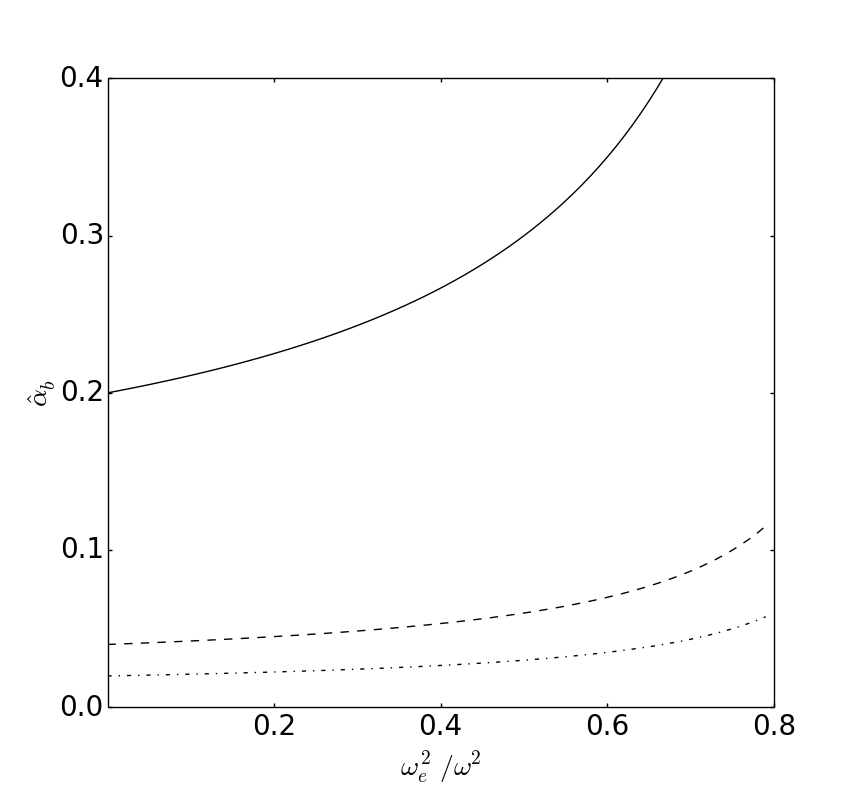}
	b.\includegraphics[scale=0.34]{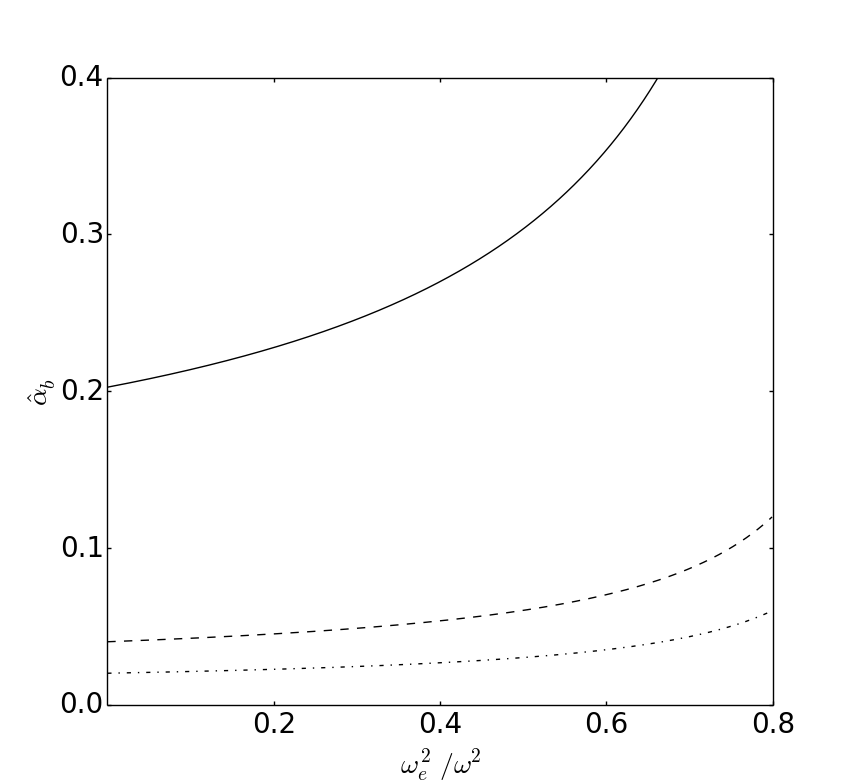}
	\caption{a) Plot of $\hat{\alpha}_b$ vs. $\omega^2_e/\omega^2$ for $b/2M=10$ (continuous line), $b/2M=50$ (dashed line), and $b/2M=100$ (dot-dashed line) for uniform plasma. b) Plot of $\hat{\alpha}_b$ vs. $\omega^2_e/\omega^2$ for the rotating case. We used different values of the impact parameter: $b/2M=10$ (continuous line), $b/2M=50$ (dashed line), and $b/2M=100$ (dot-dashed line).  We assumed $\Lambda=0.5$, $J_r/M^2=0.25$, $\sin\chi=1$, and $\omega^2_e/\omega^2=0.5$. Note that there is a small increment for $b/2M=10$ when we compare with  Schwarzschild (left panel). Figures taken from Ref.~\cite{Benavides-Gallego:2018ufb}.\label{fig1}}
\end{figure}

In Fig.~\ref{fig1} left we plotted $\hat{\alpha}_b$ as a function of $\omega^2_e/\omega^2$ for different values of $b/2M$. The plot shows that $\hat{\alpha}_b$ increases as the ration $\omega^2_e/\omega^2$ increases. On the other hand, for small values of $b/2M$ the values of the deflection angle are greater. For example, for $b/2M=100$ the figure shows that $\hat{\alpha}_b$ is greater than $0.2$; however, for $b/2M=50,100$ the deflection angle is less than $0.1$. It is also possible to see from the figure that $\hat{\alpha}_b$ has the value $4M/b$ when there is not plasma ($\omega_e=0$).

%%%%%%%%%%%%%%%%%%%%%%%%%%%%%%%%%%%%%%%%%%%%%%%%%%%%%%%%%%%%%
\subsection{Slowly rotating case}
To study the behavior of the deflection angle in the slowly rotating case, we first express the line element in Eq.~(\ref{3.1}) in the form~\cite{Morozova13,Hartle:1968si}
\begin{equation}
\label{3.2.1}
ds^2=-\left(1-\frac{2M}{r}\right)dt^2+\left(1-\frac{2M}{r}\right)^{-1}dr^2+r^2(d\theta^2+\sin^2\theta d\phi^2)-2\overline{\omega}_{LT}r^2\sin^2\theta dtd\phi.
\end{equation}

\begin{figure}[h!]
	\centering	
	\includegraphics[scale=0.15]{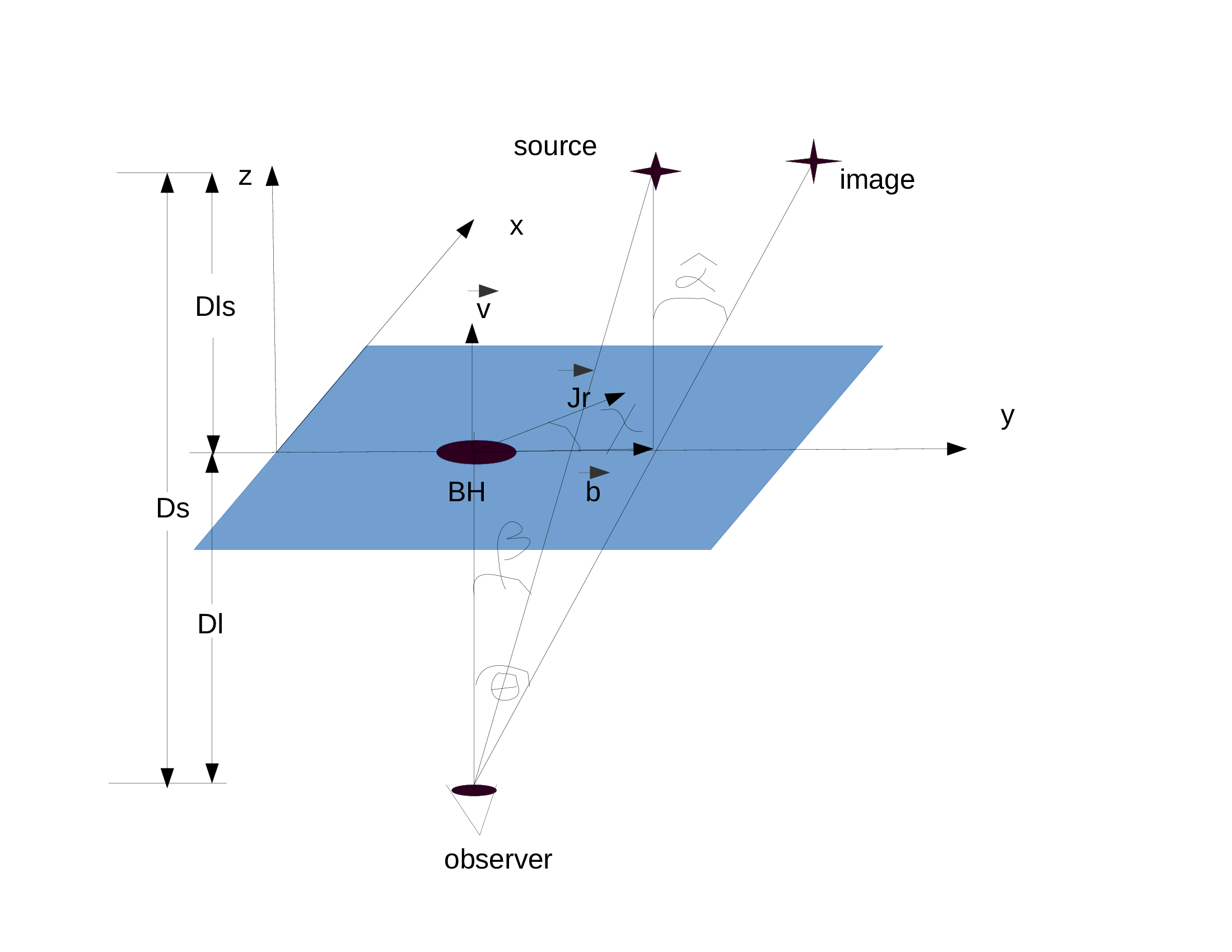}
	\caption{Schematic representation of the gravitational lensing system.  Here, $\chi$ represents the inclination angle between the vectors $\mathbf{J}_r$ and $\mathbf{b}$. In the figure, $D_s$, $D_{l}$, and $D_{ls}$ are the distances from the source to the observer, from the lens to the observer, and from the source to the lens, respectively. Figures taken from Ref.~\cite{Benavides-Gallego:2018ufb}. \label{fig2}}
\end{figure} 

where $\overline{\omega}_{LT}={2Ma}/{r^3}={2\overline{J}}/{r^3}$, with $\overline{J}={J}/{\Lambda}$, is the Lense-Thirring angular velocity of the dragging of inertial frame. In this sense, due to the presence of non-diagonal terms in the line element (\ref{3.2.1}), we use the form of $\hat{\alpha}$ obtained in Eq.~(\ref{2.2.4}). Nevertheless, in ordet to obtain the small perturbation $h_{ik}$, we recall that the dragging effect on the inertial frame contributes to $\hat{\alpha}$ only by means of the projection $\overline{J}_r$ of the angular momentum~\cite{Morozova13}. Thus, after the introduction of polar coordinates $(b,\chi)$ on the intersection point between the light ray and the $xy$-plane, where $\chi$ is the angle between $\vec{J}_{r}$ and $\vec{b}$, we find that~\cite{Morozova13}~(see Fig.~\ref{fig2}) 

\begin{equation}
\label{3.2.2}
h_{03}=-2\frac{\overline{J}_rb\sin\chi}{(b^2+z^2)^{3/2}}.
\end{equation}
Note that the deflection angle in Eq.~(\ref{2.2.4}) will contain two contributions since $h_{03}$ depends on $b$ and $\chi$\footnote{In this manuscript we only consider de case in which $\chi=\pi/2$ so that $\hat{\alpha}_\chi=0$}. These contirbutions are given by~\cite{Benavides-Gallego:2018ufb} 
\begin{equation}
\label{3.2.3}
\begin{aligned}
\hat{\alpha}_b&=\hat{\alpha}_{bS}-2\overline{J}_r\sin\chi\int^\infty_{0}\left(\frac{1}{n(b^2+z^2)^{3/2}}-\frac{3b^2}{n(b^2+z^2)^\frac{5}{2}}\right)dz\\\\
\hat{\alpha}_\chi&=-2\overline{J}_r\cos\chi\int^\infty_{0}\frac{1}{n(b^2+z^2)^{3/2}}dz ,
\end{aligned}
\end{equation}
where $\hat{\alpha}_{bS}$ is the deflection angle for Schwarzschild given by Eq.~(\ref{3.1.5}). Thus, for  ${\omega^2_e}/{\omega^2}\ll 1$, $\hat{\alpha}_b$ reduces to~\cite{Benavides-Gallego:2018ufb} 
\begin{equation}
\label{3.2.4}
\begin{aligned}
\hat{\alpha}_b&=\underbrace{\frac{4M}{b}}_{\hat{\alpha}_{S1}}+\underbrace{\frac{2Mb}{\omega^2}\int^\infty_0\frac{\omega^2_e}{r^3}dz}_{\hat{\alpha}_{S2}}+\underbrace{\frac{bK_e}{\omega^2}\int^\infty_0\frac{1}{r}\frac{dN}{dr}dz}_{\hat{\alpha}_{S3}}+\underbrace{\frac{bK_e}{\omega^4}\int^\infty_0\frac{\omega^2_e}{r}\frac{dN}{dr}dz}_{\hat{\alpha}_{S4}}\\
&+\underbrace{\frac{2J_r}{\Lambda b^2}\sin\chi}_{\hat{\alpha}_{B1}}-\underbrace{\frac{J_r}{\Lambda\omega^2}\sin\chi\int^\infty_0\frac{\omega_e^2}{r^3}dz}_{\hat{\alpha}_{B2}}+\underbrace{\frac{3b^2J_r}{\Lambda\omega^2}\sin\chi\int^\infty_0\frac{\omega_e^2}{r^5}dz}_{\hat{\alpha}_{B3}},
\end{aligned}
\end{equation}
here $r=\sqrt{b^2+z^2}$, and $S$ and $B$ stand for Schwarzschild and Boosted, respectively.  Equation~(\ref{3.2.4}) is similar to that obtained by Kogan et. al in Ref.~\cite{Kogan10}: we also find the vacuum gravitational deflection $\hat{\alpha}_{S1}$, the correction to the gravitational deflection due to the presence of the plasma $\hat{\alpha}_{S2}$, the refraction deflection due to the inhomogeneity of the plasma $\hat{\alpha}_{S3}$, and its small correction $\hat{\alpha}_{S4}$. However, when the boosted Kerr metric is considered, three more terms appear: $\hat{\alpha}_{B1}$, $\hat{\alpha}_{B2}$, and $\hat{\alpha}_{B3}$. These are contributions due to the dragging of the inertial frame. The former is a constant that is presented in all distributions considered. The other two depend on plasma distribution. 

\begin{figure}[h!]
	\centering
	a.\includegraphics[scale=0.34]{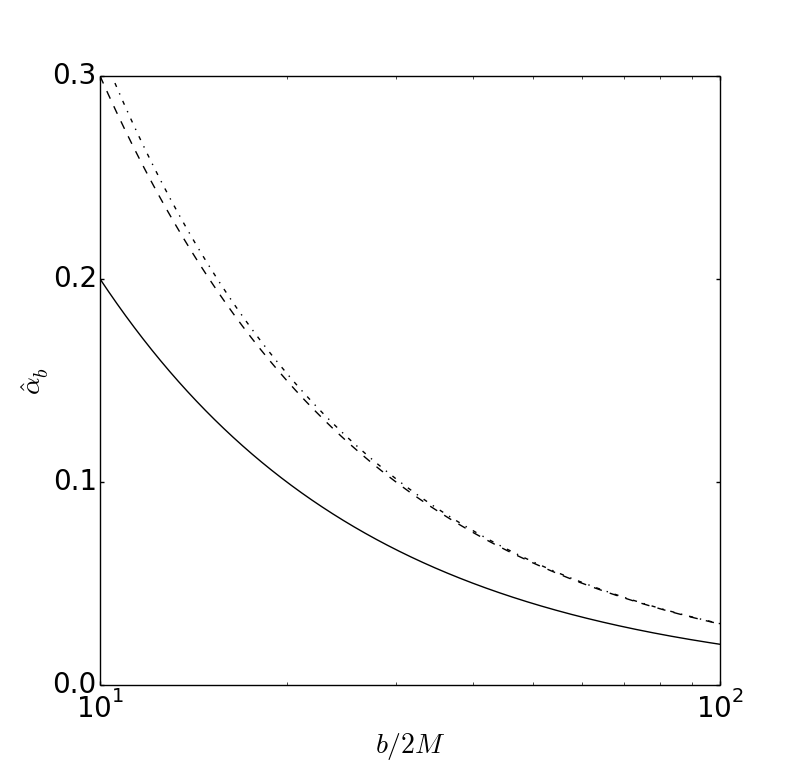}
	b.\includegraphics[scale=0.32]{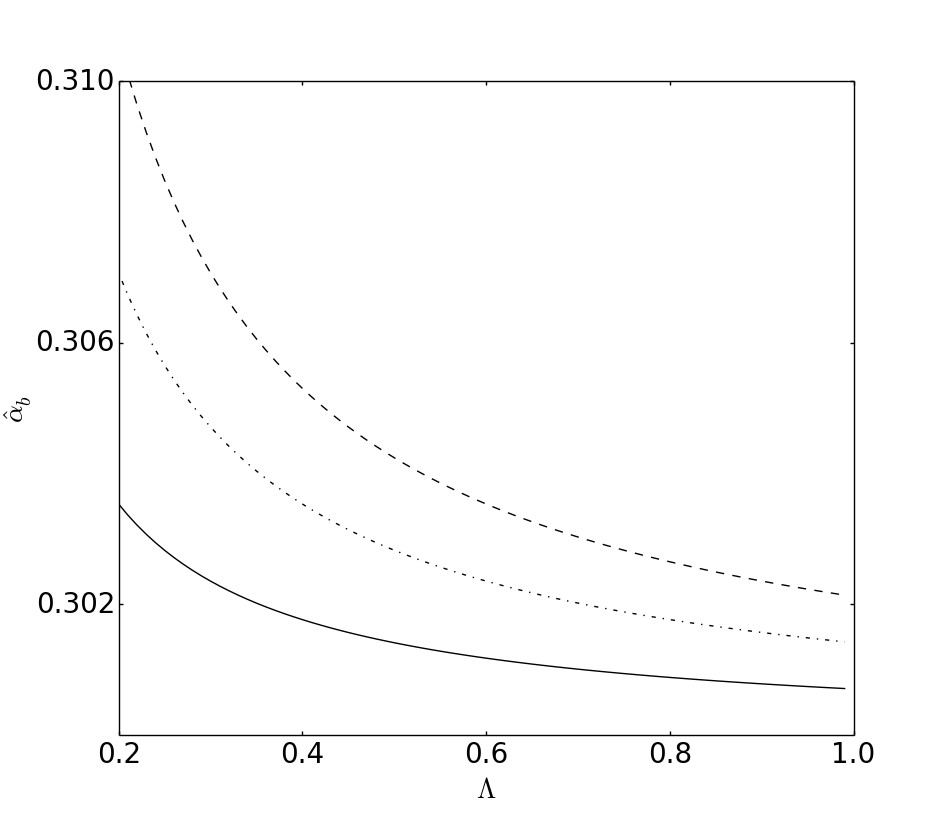}
	\label{alpha_vs_b2M_uniform_plasma_rotating_case}
	\caption{a) Plot of $\hat{\alpha}_b$ vs. $b/2M$ in the presence of uniform plasma for the slowly rotating (dot-dashed line) and $\hat{\alpha}_{bS}$ (dashed line). In the figure it is also plotted the Schwarzschild case in vacuum (continuous line). We used $\Lambda=0.5$, $J_r/M^2=0.25$, $\sin\chi=1$, and $\omega^2_e/\omega^2=0.5$. b) Plot of $\hat{\alpha}_b$ vs. $\Lambda$ for $J_r/M^2=0.1$ (continuous line), $J_r/M^2=0.2$ (dot-dashed line) , and $J_r/M^2=0.3$ (dashed line). We assumed  $b/2M=10$, $\sin\chi=1$, and $\omega^2_e/\omega^2=0.5$. Figures taken from Ref.~\cite{Benavides-Gallego:2018ufb}. \label{fig3}}
\end{figure}

In the presence of uniform plasma, the deflection angle in Eq.~(\ref{3.2.4}) takes the form
\begin{equation}
\label{3.2.5}
\hat{\alpha}_b=\underbrace{\frac{2M}{b}\left(1+\frac{1}{1-\frac{\omega^2_e}{\omega^2}}\right)}_{\hat{\alpha}_{bS}}+\underbrace{\frac{1}{\sqrt{1-\frac{\omega^2_e}{\omega^2}}}\frac{2J_r}{b^2\Lambda}}_{\hat{\alpha}_{bD}}.
\end{equation}
In Fig.~\ref{fig3} left panel, we plot $\hat{\alpha}_{bS}$ and $\hat{\alpha}_b$ for the slowly rotating case as a function of the impact parameter $b/2M$. From the figure, we see that $\hat{\alpha}_b$ for a boosted Kerr black hole is greater than $\hat{\alpha}_{bS}$. This is due to the rotation and boosts velocity $v$, which is larger for small values of $b/2M$. On the other hand, for larger values of the impact parameter $b/2M$, this difference becomes very small, and both angles behave in the same way since ${2J_r}/(nb^2\Lambda)\rightarrow 0$ when $b/2M\rightarrow\infty$.\\

On the other hand, in Fig.~\ref{fig3} right panel, we plot Eq.~(\ref{3.2.5}) as a function of $\Lambda$ for different values of $J_r$. We took into account the condition in which $0<\Lambda\leq 1$ in order to give the values. In this figure, for different values of $\Lambda$, we see that $\hat{\alpha}_b$ is bigger when $\Lambda\rightarrow 0$. Moreover, for $\Lambda=1$, the deflection angle reduces to the value $\hat{\alpha}_{bS}+{2J_r}/{nb^2}$.

Now, in order to study the behavior of $\hat{\alpha}$ in the presence of non-uniform plasma, it is necessary to know, according to Eq.~(\ref{3.2.5}), the plasma concentration $N(r)$ and plasma frequency $\omega^2_e$; both functions of the distribution density $\rho(r)$:
\begin{equation}
\label{3.2.6}
\begin{array}{ccc}
N(r)=\frac{\rho(r)}{\kappa m_p}&\text{and}&\omega^2_e=K_eN(r)=\frac{K_e\rho(r)}{\kappa m_p},
\end{array}
\end{equation}
where $m_p$ is the proton mass and $\kappa$ is is a non-dimensional coeficient related to the dark matter contribution~\cite{Kogan10}.

In the case of a singular isothermal sphere (\textbf{SIS}), the distribution density is given by 
\begin{equation}
\label{3.2.7}
\rho(r)=\frac{\sigma^2_v}{2\pi r^2},
\end{equation} 
where $\sigma^2_v$ is a one-dimensional velocity dispersion. The \textbf{SIS} model is often used in lens modeling of galaxies and clusters~\cite{ChandrasekhaBook39}. Hence, after using Eq.~(\ref{3.2.4}), the deflection angle reduces to~\cite{Benavides-Gallego:2018ufb}
\begin{equation}
\label{3.2.8}
\hat{\alpha}_{SIS}=\frac{2}{\overline{b}}+\frac{1}{12\pi}\frac{\omega^2_c}{\omega^2\overline{b}^3}-\frac{1}{16}\frac{\omega^2_c}{\omega^2\overline{b}^2}+\frac{1}{2}\frac{\widetilde{J}_r}{\Lambda \overline{b}^2}
-\frac{1}{48\pi}\frac{\widetilde{J}_r\omega^2_c}{\Lambda \omega^2\overline{b}^4}+\frac{1}{20\pi}\frac{\widetilde{J}_r\omega^2_c}{\Lambda\omega^2\overline{b}^4},
\end{equation}  
where we define $\omega^2_c=K_e\sigma^2_v/M^2\kappa m_p$, $\widetilde{J}_r=J_r/M^2$, and $\overline{b}=b/2M$. 

\begin{figure}[h!]
	\centering
	a.\includegraphics[scale=0.34]{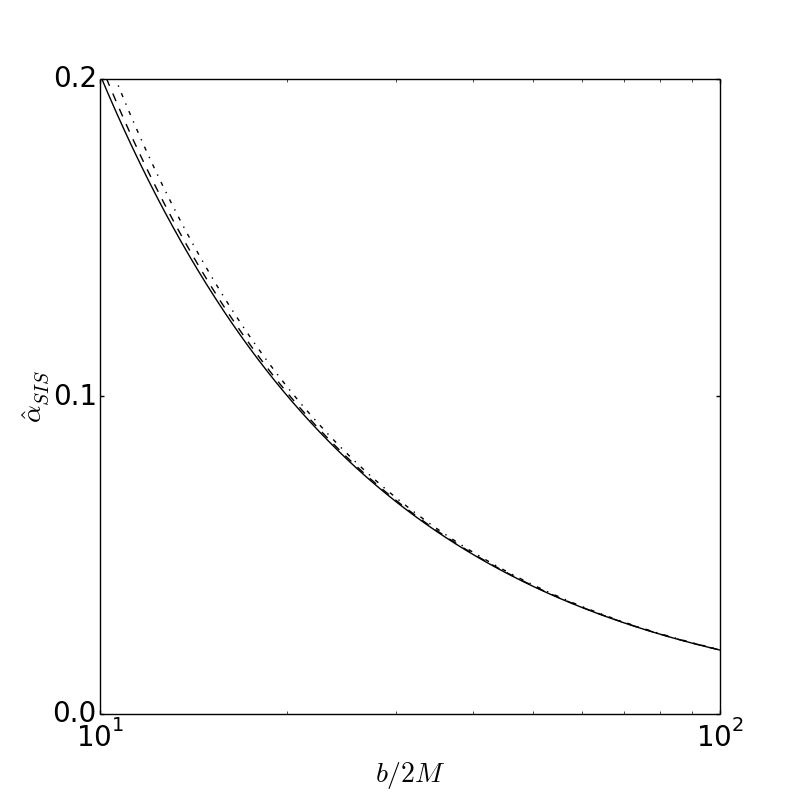}
	b.\includegraphics[scale=0.34]{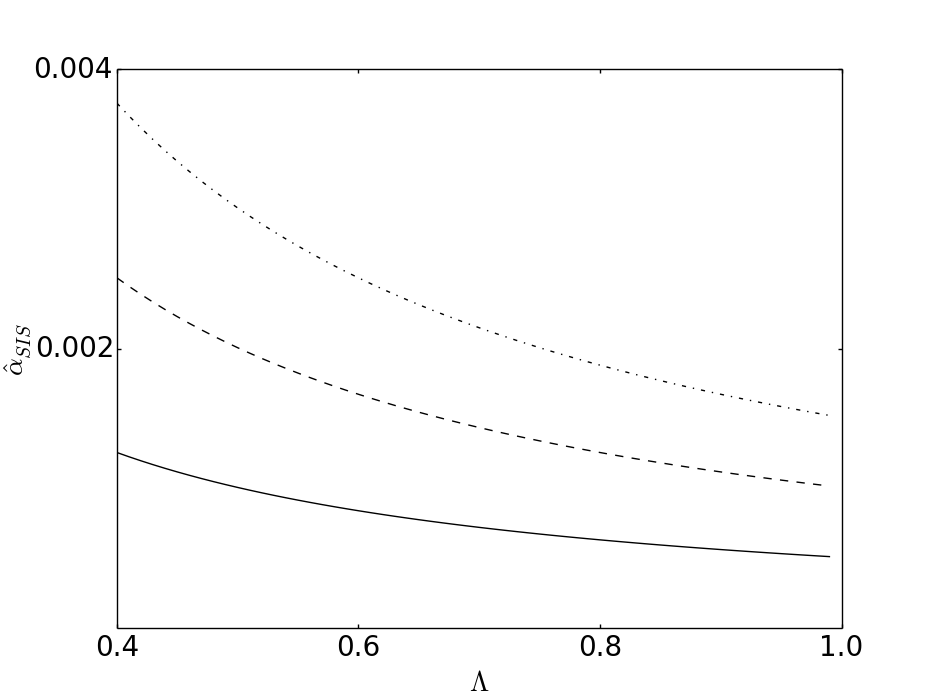}
	\caption{a) Plot of $\hat{\alpha}_{SIS}$ vs. $b/2M$ for $\Lambda=1$ (continuous line), $\Lambda=0.2$ (dashed line), and $\Lambda=0.1$ (dot-dashed line). We used $J_r/M^2=0.25$, $\sin\chi=1$, and $\omega^2_c/\omega^2=0.5$. b) Plot of $\hat{\alpha}_{SIS}$ vs. $\Lambda$ for $\widetilde{J}_r=0.1$ (continuous line), $\widetilde{J}_r=0.2$ (dashed line), and $\widetilde{J}=0.3$ (dot-dashed line). We used, $\overline{b}=10$, $\sin\chi=1$, and $\omega^2_c/\omega^2=0.5$. Figures taken from Ref.~\cite{Benavides-Gallego:2018ufb} \label{fig4}}
\end{figure}
In Fig.~\ref{fig4} left panel, we show the plot of $\hat{\alpha}_{SIS}$ as a function of $\overline{b}$ for different values of $\Lambda$. According to the figure, there is no difference for values of $b/2M$ greater than 10. However, for $b/2M$ near to 10, there is a small difference. This means that $\hat{\alpha}_{SIS}$ is greater when $\Lambda$ is small. When $\Lambda=1$ ($v=0$), we have the case of a slowly rotating massive object. In this sense, $\Lambda$ has a small effect on the deflection angle. this behavior can be seen clearly in the right panel of Fig.~\ref{fig4}, where it is plotted $\hat{\alpha}_{SIS}$ as a function of $\Lambda$ for different values of $\widetilde{J}_r$. Note that the boosted parameter is constrained to the interval $0<\Lambda\leq 1$. 

\begin{figure}[h!]
	\centering
	a.\includegraphics[scale=0.34]{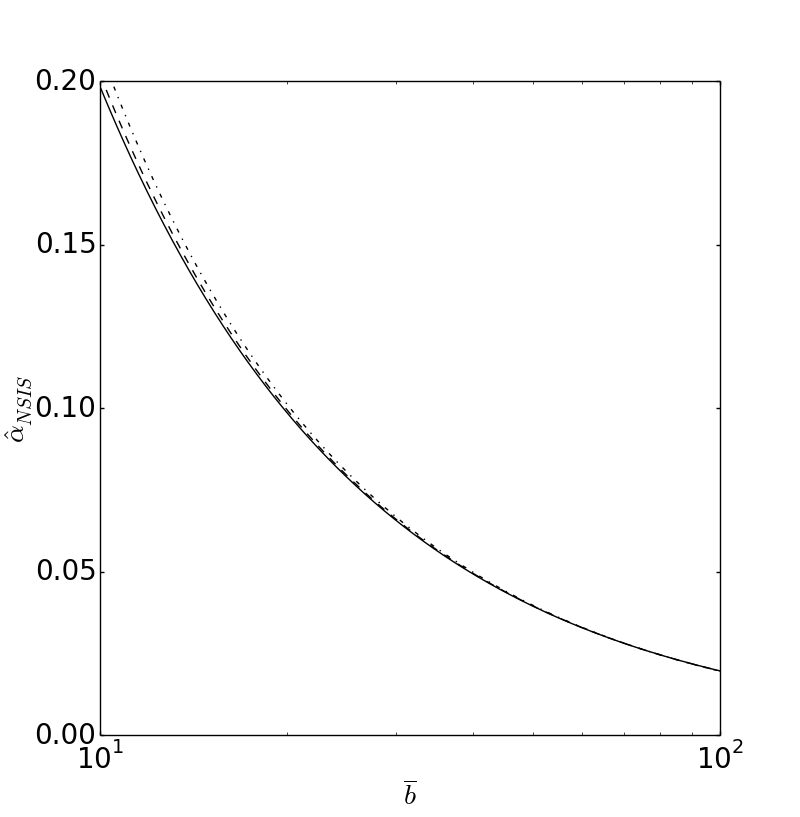}
	b.\includegraphics[scale=0.34]{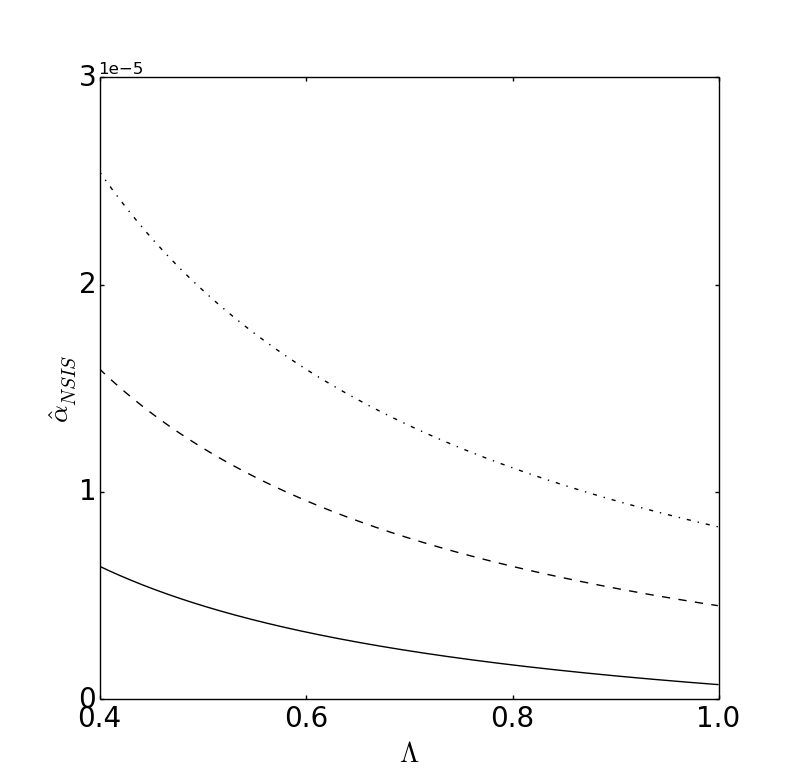}
	\caption{a) Plot of $\hat{\alpha}_{NSIS}$ vs. $\overline{b}$ for $\Lambda=1$ (continuous line), $\Lambda=0.25$ (dashed line), and $\Lambda=0.1$ (dot-dashed line). We used, $\widetilde{J}_r=0.25$, $\overline{r}_c=10$, $\sin\chi=1$, and $\omega^2_c/\omega^2=0.5$. b) Plot of $\hat{\alpha}_{NSIS}$ vs. $\Lambda$ for $\widetilde{J}_r=0.1$ (continuous line), $\widetilde{J}_r=0.2$ (dashed line), and $\widetilde{J}=0.3$ (dot-dashed line). We used, $\overline{b}=100$, $\overline{r}_c=10$, $\sin\chi=1$, and $\omega^2_c/\omega^2=0.5$. Note the scale used for the deflection angle: each value is multiplied by $1e-5=1\times 10^{-5}$. Figures taken from Ref.~\cite{Benavides-Gallego:2018ufb}. \label{fig5}}
\end{figure}

The non-singular isothermal sphere (\textbf{NSIS}) is a model of plasma distribution where the singularity is replaced by a finite core. In this plasma, the density distribution is given by~\cite{Hinshaw87}
\begin{equation}
\label{3.2.9}
\begin{array}{ccc}
\rho(r)=\frac{\sigma^2_v}{2\pi(r^2+r^2_c)}=\frac{\rho_0}{\left(1+\frac{r^2}{r^2_c}\right)}&\text{with}&\rho_0=\frac{\sigma^2_v}{2\pi r^2_c}.
\end{array}
\end{equation} 
Here $r_c$ is the core radius. In the presence of a \textbf{NSIS}, the deflection angle in Eq.~(\ref{3.2.5}) takes the form~\cite{Benavides-Gallego:2018ufb}
\begin{equation}
\label{3.2.10}
\begin{aligned}
\hat{\alpha}_{NSIS}&=\frac{2}{\overline{b}}+\frac{2\overline{b}\omega^2_c}{\pi\omega^2}\bigg[\frac{1}{4\overline{b}^2\overline{r}^2_c}-\frac{{\rm arctanh}\bigg(\frac{\overline{r}_c}{\sqrt{4\overline{b}^2+\overline{r}^2_c}}\bigg)}{\overline{r}^3_c\sqrt{\overline{r}^2_c+4\overline{b}^2}}\bigg] -\frac{1}{2}\frac{\overline{b}\omega^2_c}{(4\overline{b}^2+\overline{r}^2_c)^\frac{3}{2}\omega^2}+\frac{2\tilde{J}_r}{\Lambda \overline{b}^2}\\
&-\frac{\widetilde{J}_r\omega^2_c}{2\pi\Lambda\omega^2}\left[\frac{1}{4\overline{b}^2\overline{r}^2_c}-\frac{{\rm arctanh}\left(\frac{\overline{r}_c}{\sqrt{4\overline{b}^2+\overline{r}^2_c}}\right)}{\overline{r}^3_c\sqrt{\overline{r}^2_c+4\overline{b}^2}}\right]+\frac{6}{\pi}\frac{\overline{b}^2\widetilde{J}_r\omega^2_c}{\Lambda\omega^2}\left[\frac{2\overline{r}^2_c-12\overline{b}^2}{48\overline{b}^4\overline{r}^4_c}+\frac{{\rm arctanh}\left(\frac{\overline{r}_c}{\sqrt{4\overline{b}^2+\overline{r}^2_c}}\right)}{\overline{r}^5_c\sqrt{\overline{r}^2_c+4\overline{b}^2}}\right].
\end{aligned}
\end{equation} 
In Fig.~\ref{fig5} left panel,  we show the behaviour of $\hat{\alpha}_{NSIS}$ as a function of $\overline{b}$ for different values of $\Lambda$. In the plot, since  we are in the weak field limit, we consider $\overline{b} \gg \overline{r}_c$. The behavior is very similar to the singular plasma distribution: there are small differences in $\hat{\alpha}_{NSIS}$ when small values of $\Lambda$ are considered, and no difference appears when the impact parameter $\overline{b}$ takes values greater than $10$. Fig.~\ref{fig5}  right panel shows clearly this behavior.
Finally, we consider the deflection angle in the case of plasma in a galaxy cluster. Due to the large temperature of electrons in the cluster, the distribution can be considered as homogeneous. In this sense, it is possible to suppose a \textbf{SIS} distribution. Hence, the plasma density has the form~\cite{Kogan10}
\begin{equation}
\label{3.2.11}
\rho(r)=\rho_0\left(\frac{r}{r_0}\right)^{-s},s=\frac{2\sigma^2_v}{\mathfrak{R}T}.
\end{equation}   
Therefore, after using Eq.~(\ref{3.2.4}), the deflection angle has the form~\cite{Benavides-Gallego:2018ufb}
\begin{equation}
\label{3.2.12}
\begin{aligned}
\hat{\alpha}_{PGC}&=\frac{2}{\overline{b}}+\frac{\sqrt{\pi}}{2^{s+1}(s+1)}\frac{\overline{r}^s_0\omega^2_f}{\overline{b}^2\omega^2}\frac{\Gamma(\frac{s}{2}+1)}{\Gamma(\frac{s+1}{2})}-\frac{\sqrt{\pi}}{2^s}\frac{\omega^2_f}{\omega^2}\frac{\Gamma(\frac{s}{2}+1)}{\Gamma(\frac{s}{2})}\left(\frac{\overline{r}_0}{\overline{b}}\right)^s+\frac{2\tilde{J}_r}{\Lambda \overline{b}^2}\\
&-\frac{\pi}{2^{s+2}(s+1)}\frac{\widetilde{J}_r\overline{r}^2_0\omega^2_f}{\overline{b}^{s+2}\Lambda\omega^2}\frac{\Gamma(\frac{s}{2}+1)}{\Gamma(\frac{s+1}{2})}+\frac{3\sqrt{\pi}}{2^{s+2}(s+3)}\frac{\widetilde{J}_r \overline{r}^s_0\omega^2_f}{b^{s+2}\Lambda\omega^2}\frac{\Gamma(\frac{s+4}{2})}{\Gamma(\frac{s+1}{2})}.
\end{aligned}
\end{equation}
where we define $\omega^2_f=\frac{K_e\rho_0}{\kappa m_p}$, $\overline{r}_0=r_0/M$, $\widetilde{J}_r=J_r/M^2$, and $\overline{b}=b/2M$. 

\begin{figure}[h!]
	\centering
	a.\includegraphics[scale=0.22]{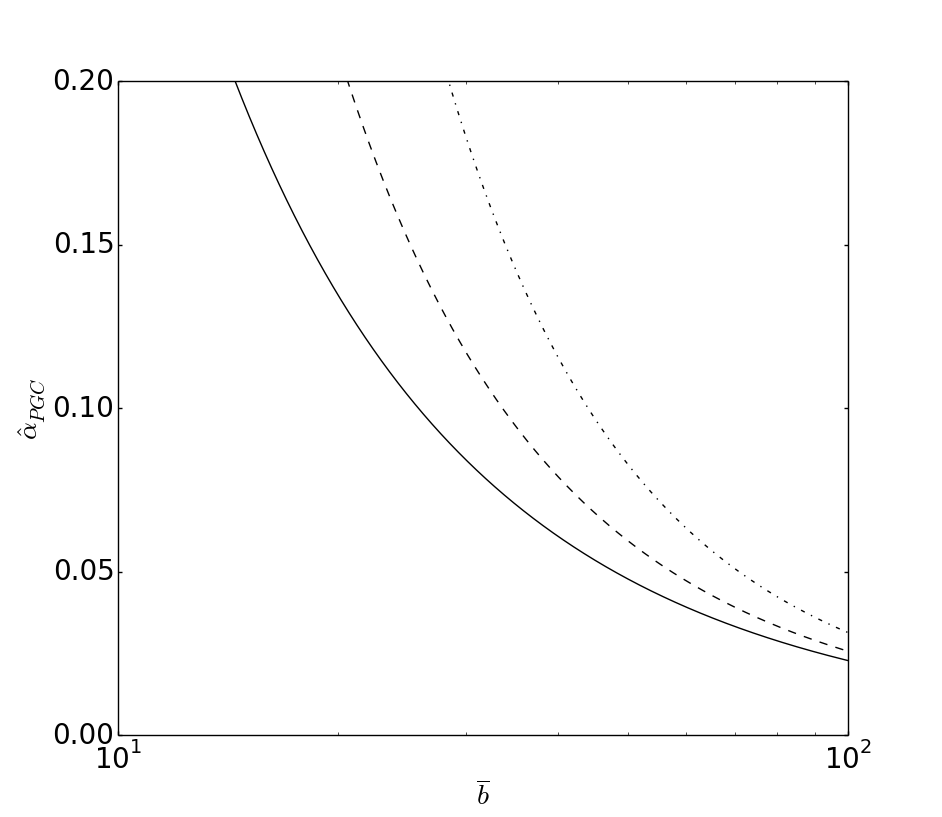}
	b.\includegraphics[scale=0.22]{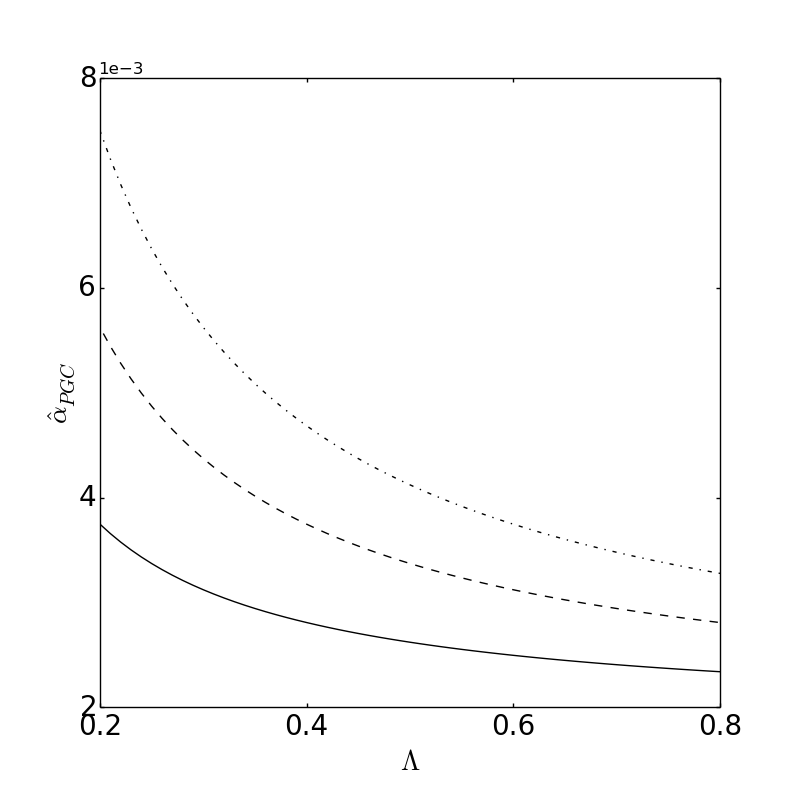}
	c.\includegraphics[scale=0.22]{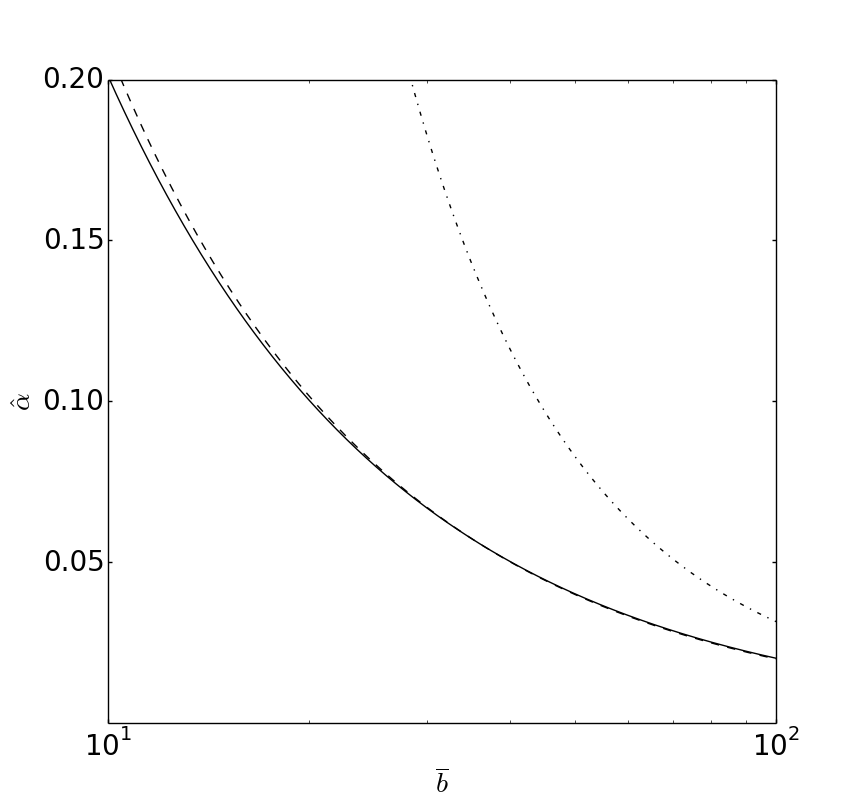}
	\caption{a) Plot of $\hat{\alpha}_{PGC}$ vs. $\overline{b}$ for $\Lambda=1$ (continuous line), $\Lambda=0.25$ (dashed line), and $\Lambda=0.1$ (dot-dashed line). We used, $\widetilde{J}_r=0.25$, $\overline{r}_0=10$, $\sin\chi=1$, $s=0.03$, and $\omega^2_f/\omega^2=0.5$. b) Plot of $\hat{\alpha}_{PGC}$ vs. $\Lambda$ for $\widetilde{J}_r=0.1$ (continuous line), $\widetilde{J}_r=0.2$ (dashed line), and $\widetilde{J}_r=0.3$ (dot-dashed line). We used $\overline{r}_0=10$, $\sin\chi=1$, $s=0.03$, $\overline{b}=100$ and $\omega^2_f/\omega^2=0.5$. Note the scale used for the deflection angle: each value is multiplied by $1e-3=1\times 10^{-3}$. c) Plot of $\hat{\alpha}$ vs. $\overline{b}$ for  \textbf{SIS} (continuous line), \textbf{NSIS} (dashed line), and \textbf{PGC} (dot-dashed line). We used $\Lambda=0.1$, $\overline{r}_c=10$, $\overline{r}_0=10$, $\sin\chi=1$, $s=0.03$, and $\omega^2_f/\omega^2=\omega^2_c/\omega^2=0.5$. For \textbf{NSIS} we use $\Lambda=1$ since no difference from \textbf{SIS} was found. Figures taken from Ref.~\cite{Benavides-Gallego:2018ufb}.\label{fig6}}
\end{figure}

In Figs.~\ref{fig6} a, b we plot $\hat{\alpha}_{PGC}$ as a function of $\overline{b}$ and $\Lambda$, respectively. In In order to obtain these plots we considered the case $s<<1$ \cite{Kogan10}. According to Figs.~\ref{fig6}, differences in the deflection angle can be seen clearly for the \textbf{PGC} distribution when compared with the previous distributions. Furthermore, Fig.~\ref{fig6} b shows that the deflection angle increases due to the dragging and small values of $\Lambda$. In Fig.~\ref{fig6} c, we show the behavior of $\hat{\alpha}$ as a function of the impact parameter $\overline{b}$ for all distributions. Observ that values of $\hat{\alpha}$ for the \textbf{PGC} distribution are grater than the other two distributions. In the figure, it is also possibel to see the small difference between \textbf{SIS} and \textbf{NSIS} distributions for small values of $b/2M$.\\ 

%%%%%%%%%%%%%%%%%%%%%%%%%%%%%%%%%%%%%%%%%%%%%%%%%%%%%%%%%%%%%
\section{\label{sec:magnification} Lens equation and magnification in the presence of plasma:}

In this section, as an application, we study the magnification for boosted Kerr black hole in the presence of plasma. In particular, we compare the uniform and singular isothermal distributions. 

It is known that the magnification of brightness of the star is defined as~\cite{Morozova13} 
\begin{equation}
\label{4.1}
\begin{array}{cc}
\mu_\Sigma=\frac{I_{tot}}{I_{\ast}}=\sum_k \left|\left(\frac{\theta_k}{\beta}\right)\left(\frac{d\theta_k}{d\beta}\right)\right|,&k=1,2,...,m,
\end{array}
\end{equation}
where $m$ is the number of images, $I_{tot}$ and $I_{\ast}$ are the total and unlensed brightness of the source respectively, $\theta_k$ is the image position, and $\beta$ is the angular position of the source (see figure \ref{fig2}). Thus, to compute the magnification for different distributions, it is necessary to solve the lens equation
\begin{equation}
\label{4.2}
\theta D_s=\beta D_s+\hat{\alpha}D_{ls}.
\end{equation}
The lens Eq.~(\ref{4.2}) relates the distance from the observer to the source $D_s$ with the distance from the lens to the source $D_{ls}$. In the last equaiton, $\hat{\alpha}$ is the deflection angle, and $\theta$, $\beta$ the image and source positions respectively (see figure \ref{fig2}). In the case of uniform plasma, Eq.~(\ref{4.2}) reduces to~\cite{Benavides-Gallego:2018ufb}
\begin{equation}
\label{4.3}
\theta^3-\beta\theta^2-\frac{\theta^2_E}{2}\left(1+\frac{1}{1-\frac{\omega^2_e}{\omega^2}}\right)\theta-\frac{\theta^2_E\tilde{J}_r}{4\overline{D}_l\Lambda}\frac{1}{\sqrt{1-\frac{\omega^2_e}{\omega^2}}}=0.
\end{equation}
In the case of the \textbf{SIS}, the lens equation takes the form~\cite{Benavides-Gallego:2018ufb}
\begin{equation}
\label{4.4}
\theta^3-\beta\theta^2-\frac{2D_{ls}}{D_lD_s}\theta-\frac{\overline{D}_{ls}}{\overline{D}^2_l\overline{D}_s}\left(\frac{\tilde{J}_r}{2\Lambda}-\frac{\omega^2_c}{16\omega^2}\right)=0. 
\end{equation}
In both cases, we have considered small angles. This means that the impact parameter can be expressed as $b\approx D_l\theta$, with $D_l$ denoting the distance from the observer to the lens. 

\begin{figure}[h!]
	\centering
	a.\includegraphics[scale=0.2]{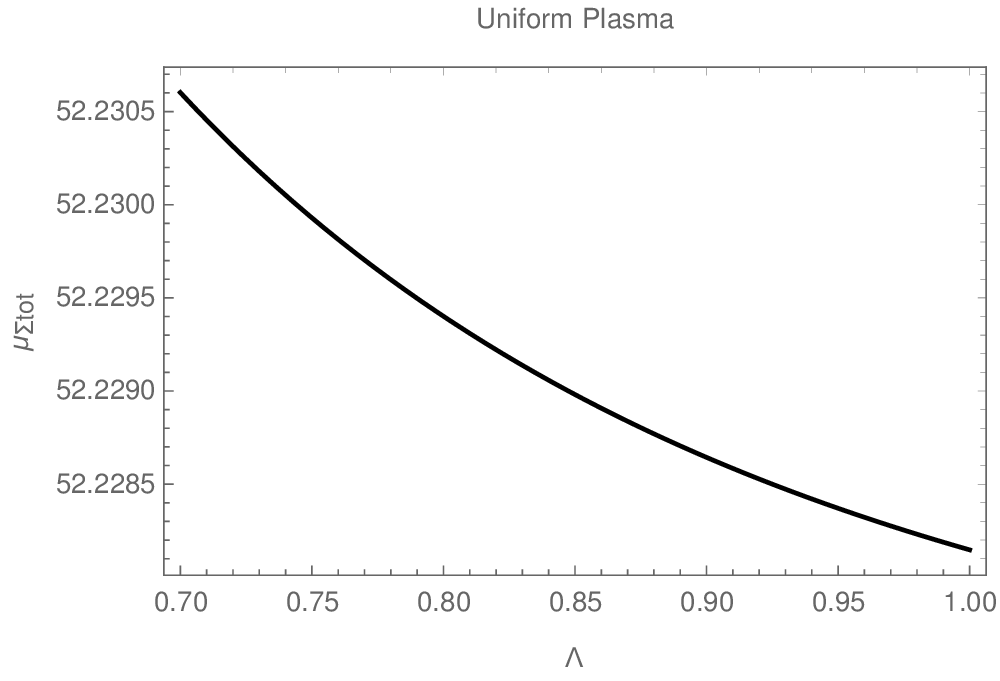}
	b.\includegraphics[scale=0.2]{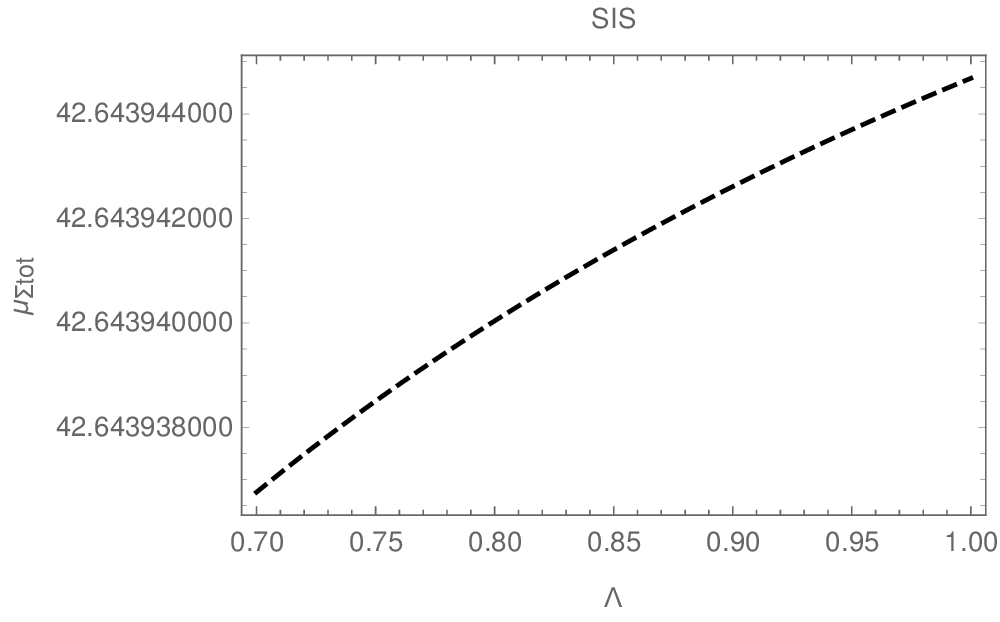}
	c.\includegraphics[scale=0.2]{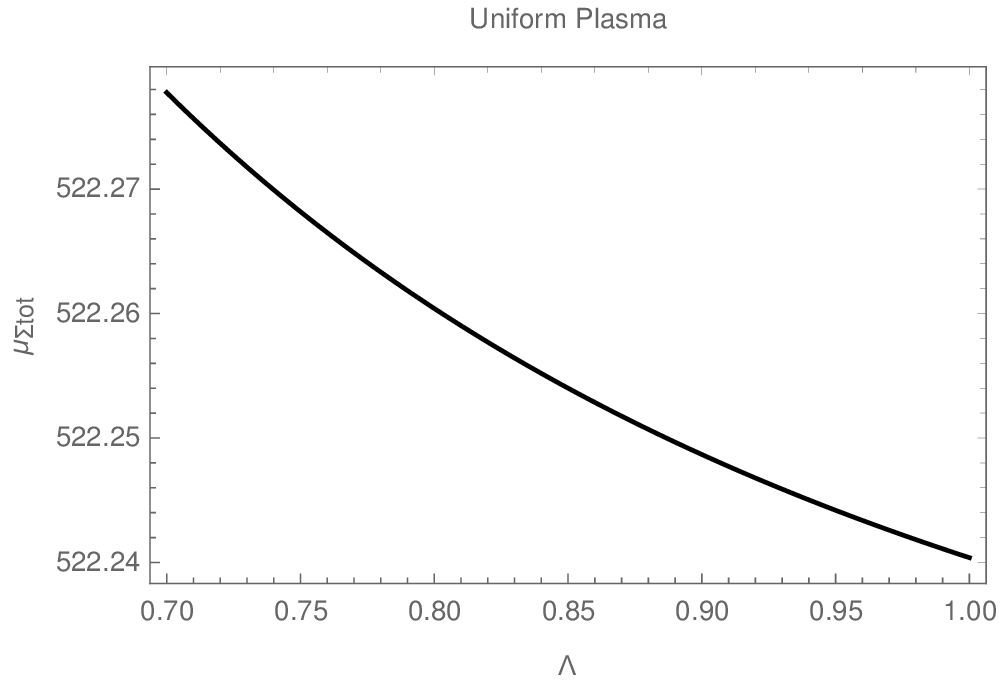}
	d.\includegraphics[scale=0.2]{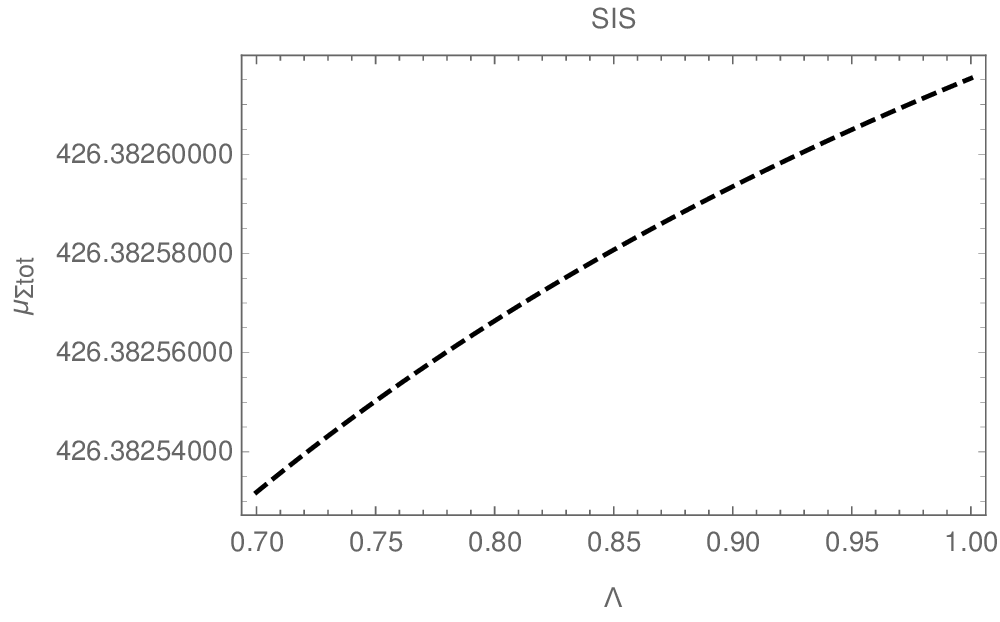}
	\caption{a) Plot of $\mu_{\Sigma tot}$ vs. $\Lambda$ when $\beta=0.001$ for uniform plasma. b) Plot of $\mu_{\Sigma tot}$ vs. $\Lambda$ when $\beta=0.001$ for the \textbf{SIS} distribution. c) Plot of $\mu_{\Sigma tot}$ vs. $\Lambda$ when $\beta=0.0001$ for unifomr plasma. (\textbf{d}) Plot of $\mu_{\Sigma tot}$ vs. $\Lambda$ when $\beta=0.0001$ for the \textbf{SIS} distribution. In all the figures we considered $\overline{D}_{ls}=10$, $\overline{D}_l=100$, $\overline{D}_s=110$, $\omega^2_e/\omega^2=\omega^2_c/\omega^2=0.5$, $\theta_E=0.001818$, and $\tilde{J}_r=0.3$. Figures taken from Ref.~\cite{Benavides-Gallego:2018ufb}. \label{fig7}}
\end{figure}

In Figs. \ref{fig7}\textcolor{blue}{.a} and \ref{fig7}\textcolor{blue}{.c}, we plotted the behaviour of the total magnification as a function of the boosted parameter $\Lambda$ for $\beta=0.001$ and $\beta=0.0001$ respectively. According to Fig. \ref{fig7}\textcolor{blue}{.a}, when $\beta=0.001$, the total magnification decreases as $\Lambda$ increases. This means that $\mu_{\Sigma tot}$ decreases as the boosted velocity $v$ of the black hole decreases. A similar behaviour can be seen from Fig. \ref{fig7}\textcolor{blue}{.c} when $\beta=0.0001$. Note that for small values of $\beta$, the magnitude of the total magnification increases. For example: when $\beta=0.001$ the total magnification is about $\mu_{\Sigma tot}\approx 52.2$. However, when $\beta=0.0001$, the value increases to $\mu_{\Sigma tot}\approx 522.2$. 

\begin{figure}[h!]
	\centering
	a.\includegraphics[scale=0.2]{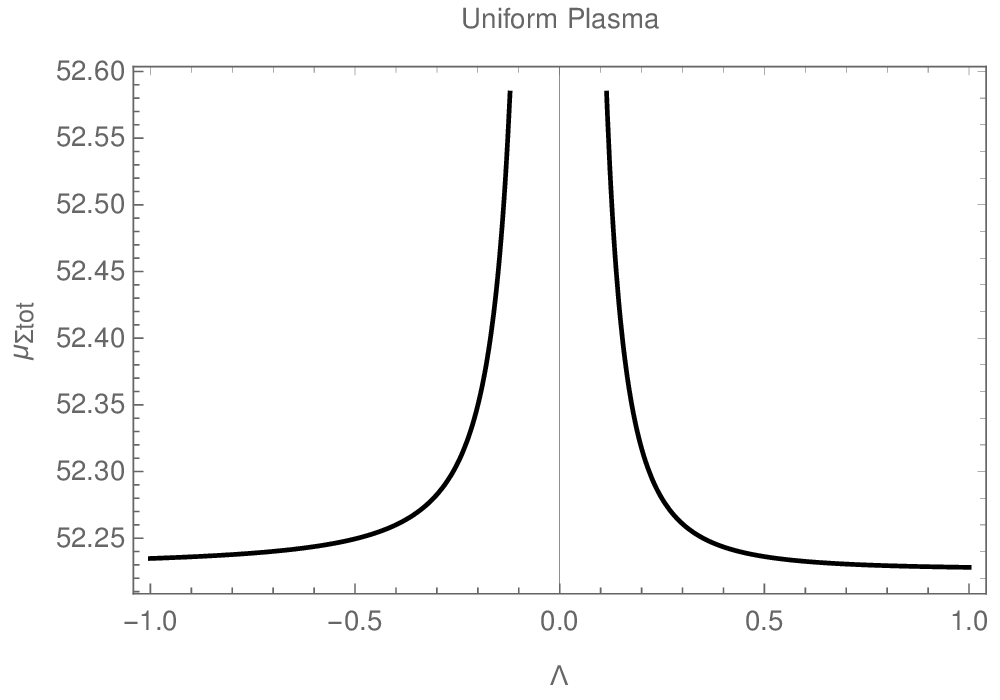}
	b.\includegraphics[scale=0.2]{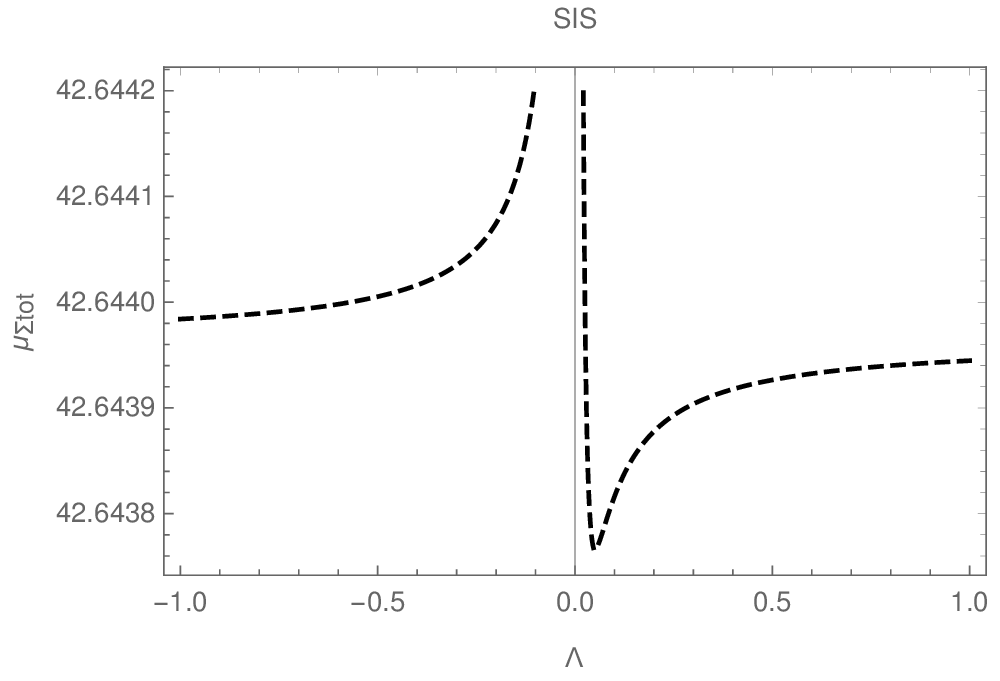}
	\caption{a) Plot of $\mu_{\Sigma tot}$ vs. $\Lambda$ when $\beta=0.001$ for uniform plasma. b) Plot of $\mu_{\Sigma tot}$ vs. $\Lambda$ when $\beta=0.001$ for the \textbf{SIS} distribution. In all the figures we considered $\overline{D}_{ls}=10$, $\overline{D}_l=100$, $\overline{D}_s=110$, $\omega^2_e/\omega^2=\omega^2_c/\omega^2=0.5$, $\theta_E=0.001818$, and $\tilde{J}_r=0.3$. Figures taken from Ref.~\cite{Benavides-Gallego:2018ufb}. \label{fig8}}
\end{figure}
On the other hand, In Figs. \ref{fig7}\textcolor{blue}{.b} and \ref{fig7}\textcolor{blue}{.d}, we plotted the behaviour of $\mu_{\Sigma tot}$ as a function of the boosted parameter $\Lambda$ for $\beta=0.001$ and $\beta=0.0001$ respectively. In contrast to uniform plasma, the total magnification increases as $\Lambda$ increases. Moreover, for small values of $\beta$, $\mu_{\Sigma tot}$ increases. Finally, in Fig.~\ref{fig8} we show the behaviour of  $\mu_{\Sigma tot}$  as a function of the boosted parameter $\Lambda$ for uniform plasma and \textbf{SIS} distributions. Note that the behaviour of $\mu_{\Sigma tot}$ in the \textbf{SIS} distribution is not symmetric.  

%%%%%%%%%%%%%%%%%%%%%%%%%%%%%%%%%%%%%%%%%%%%%%%%%%%%%%%%%%%%%

\section{Conclusions}
In this work we studied the deflection angle for a boosted Kerr metric in the presence of both uniform and non-uniform plasma distributions  (where three different cases were considered). First we studied the deflection angle for the non-rotating case in the presence of uniform plasma ($\omega_e= \text{costant}$) by considering small values of $v$. We found that $\hat{\alpha}_b$  does not dependent, at first order, on the velocity $v$. It was also found that, after the approximation $1-n\ll \frac{\omega_e}{\omega}$, the deflection angle reduces to the same expression obtained in Ref.~\cite{Kogan10} (see Eq.~\ref{3.1.6}). In the case of the slowly rotating case, the deflection angle $\hat{\alpha}_b$ in Eq.~(\ref{3.2.5}) contains two terms: the Schwarzschild angle $\hat{\alpha}_{bS}$, and the contribution due to the dragging ${\hat{\alpha}_{bD}}$. This result is quite similar to that of V.S Morozova et al. However, in contrast with their result, Eq.~(\ref{3.2.5}) also depends on the parameter $\Lambda$. Therefore, $\hat{\alpha}$ depends on $v$ only when the dragging takes place.

In the presence of non-uniform plasma, we consider the deflection angle as a function of $\overline{b}$ and $\Lambda$ for different distributions. We found that $\hat{\alpha}$ is affected by the presence of plasma and is greater when compared with vacuum and uniform distributions. Moreover, we found again that $\hat{\alpha}$ increases not only due to the dragging, but also when small values of the boosted parameter $\Lambda$ are considered. In this work, we also found two important constraints. In the case of \textbf{NSIS}, $r_c$ must have values greater than $6M$. If the core radius $r_c$ is smaller than this limit the deflection angle becomes negative at some point and will not agree with the usual behavior when $b\rightarrow\infty$~\cite{Benavides-Gallego:2018ufb}. On the other hand, regarding the \textbf{PGC}, we found that $s$ must be different from $-1$ or $-3$ as can be seen from Eq.~(\ref{3.2.12}). Nevertheless, this condition is fulfilled since we consider positive values of $s<<1$.

Finally, we compare the total magnification for uniform and \textbf{SIS} plasma distributions. According to Fig. \ref{fig7}, for small values of $v$ ($0.7\leq\Lambda\leq1$), the total magnification is greater in the case of homogeneous plasma. Furthermore, it is important to point out that the total magnification has small changes in both distributions. In the case of uniform plasma $\mu_{\Sigma tot}$ ranges from $52.2285$ to $52.2305$, and from $42.643938$ to $42.643944$ in the \textbf{SIS}. On the other hand, when we compare both models, we see that the behavior of $\mu_{\Sigma tot}$ is different. When the boosted Kerr black hole moves towards ($\Lambda>0$) or away ($\Lambda<0$) from the observer the behavior is very similar (there is a small difference when $\Lambda\rightarrow -1$ and $\Lambda\rightarrow 1$). However, when we consider the \textbf{SIS} distribution, the behavior is not symmetric. In both cases, this behavior is due to cinematic effects.

%%%%%%%%%%%%%%%%%%%%%%%%%%%%%%%%% Acknowledge %%%%%%%%%%%%%%%%%%%%
\begin{acknowledgments}
	C.A.B.G. acknowledges support from the China Scholarship Council (CSC), grant No.~2017GXZ019022. 
\end{acknowledgments}

%%%%%%%%%%%%%%%%%%%%%%%%%%%%%%%%%%%%%%%%%%%%%%%%%%%%%%%%%%%%%

\end{document}